\documentclass[12pt]{iopart}

\usepackage[dvips]{graphicx}
\usepackage[usenames,dvipsnames]{xcolor} 
\usepackage[latin1]{inputenc}
\usepackage[T1]{fontenc}
\usepackage{latexsym}
\usepackage{amsfonts}
\usepackage{amssymb}
\usepackage{amsthm}
\expandafter\let\csname equation*\endcsname\relax 
\expandafter\let\csname endequation*\endcsname\relax
\usepackage{amsmath}
\usepackage{color}
\usepackage{psfrag}
\usepackage{rotating}
\usepackage{pgf,pgfsys,pgffor}
\usepackage{pgfplots}
\usepackage{pgfplotstable}
\usepackage[caption=false]{subfig}
\usepackage{tikz}
\usepackage{hyperref}
\renewcommand{\tr}[2][]{\textrm{Tr}_{#1} \left[ {#2} \right]} 
\newcommand{\bra}[1]{\langle {#1} \vert}
\newcommand{\ket}[1]{\vert {#1} \rangle}
\newcommand{\pure}[1]{\vert {#1} \rangle \langle {#1} \vert}

\newcommand{\ave}[1]{\left \langle {#1} \right \rangle}						

\newcommand{\beq}{\begin{equation}}
\newcommand{\eeq}{\end{equation}}

\newcommand{\ito}{It\={o} }

\newcommand{\expterm}{x}				
\newcommand{\Prob}[1]{\wp \left( #1 \right)}
\newcommand{\ost}{\textrm{ost}}
\newcommand{\ProbOst}[1]{\wp_{\ost} \left( #1 \right)}

\renewcommand{\d}{\textrm{d}}		
\newcommand{\dt}{\textrm{d}t}		
\newcommand{\ds}{\textrm{d}s}		
\newcommand{\dz}{\textrm{d}z}		
\newcommand{\dv}{\textrm{d}v}		
\newcommand{\dW}{\textrm{d}W}		

\newcommand{\X}{\Phi}
\newcommand{\Xt}{\X_t}
\newcommand{\Xs}{\X_s}

\newcommand{\Xu}{\X_z}
\newcommand{\Xv}{\X_v}

\newcommand{\E}{\xi}
\newcommand{\Et}{\E_t}
\newcommand{\Es}{\E_s}
\newcommand{\Ez}{\E_z}
\newcommand{\Ev}{\E_v}

\newcommand{\J}{J}

\newcommand{\R}{R}
\renewcommand{\S}{S}

\renewcommand{\u}{u}
\newcommand{\U}{U}
\newcommand{\ut}{\u_t}
\newcommand{\us}{\u_s}
\newcommand{\uv}{\u_v}

\newcommand{\y}{\lambda}

\newcommand{\ys}{\y_s}
\newcommand{\yv}{\y_v}
\newcommand{\yu}{\y_z}
\newcommand{\ytau}{\y_\tau}
\newcommand{\qubit}{\psi}
\newcommand{\delay}{\tau}
\newcommand{\dD}{\d \Delta}
\newcommand{\dd}{\d \delta}
\newcommand{\utD}{\u_{t-\Delta}}

\newcommand{\utDd}{\u_{t-\Delta-\delta}}
\newcommand{\XtD}{\X_{t-\Delta}}

\newcommand{\XtDd}{\X_{t-\Delta-\delta}}
\newcommand{\EtD}{\E_{t-\Delta}}
\newcommand{\Etd}{\E_{t-\delta}}
\newcommand{\ytD}{\y_{t-\Delta}}
\newcommand{\ytDd}{\y_{t-\Delta-\delta}}
\newcommand{\w}{w}

\newcommand{\merit}{F}
\newcommand{\am}{\tilde{F}}

\newcommand{\tl}{t_1} 

\newcommand{\erf}[1]{Eq.~(\ref{#1})}

\newcommand{\dg}{^\dagger}

\newcommand{\xfrac}[2]{{#1}/{#2}}

\newcommand{\rnfrac}[2]{({#1})/{#2}}

\newcommand{\braket}[2]{\langle{#1}|{#2}\rangle}

\newcommand{\sq}[1]{\left[ {#1} \right]}

\newcommand{\ro}[1]{\left( {#1} \right)}

\newcommand{\st}[1]{\left| {#1} \right|}

\definecolor{cyan}{gray}{0.75}
\definecolor{magenta}{gray}{0.63}
\definecolor{green}{gray}{0.43}
\definecolor{blue}{gray}{0.3}

\definecolor{nblue}{rgb}{0.3,0.3,1.0}
\definecolor{ngreen}{rgb}{0.2,0.7,0.2}
\definecolor{nred}{rgb}{0.9,0.1,0}

\newcommand{\blk}{\color{black}}
\colorlet{NGREEN}{ngreen} 
\colorlet{NBLUE}{nblue}
\colorlet{VIOLET}{violet}
\colorlet{NRED}{nred}
\colorlet{BLACK}{black}

\begin{document}

\title[Deterministic preparation of superpositions of vacuum plus one photon]{Deterministic preparation of superpositions of vacuum plus one photon by adaptive homodyne detection: experimental considerations}

\author{Nicola Dalla Pozza$^{1,2,3}$, Howard M. Wiseman$^2$, and Elanor H. Huntington$^3$}
  \address{$^1$Department of Information Engineering (DEI), University of Padova, Padova, Italy}
  \address{$^2$Centre for Quantum Computation and Communication Technology (Australian Research Council), Centre for Quantum Dynamics, Griffith University, Brisbane, Queensland 4111, Australia}
  \address{$^3$Centre for Quantum Computation and Communication Technology, School of Engineering and Information Technology, University of New South Wales, Canberra  ACT 2600, Australia}%
\ead{n.dallapozza@unsw.edu.au,\ nicola.dallapozza@dei.unipd.it}
\vspace{10pt}
\begin{indented}
\item[]\today
\end{indented}

\begin{abstract}
The preparation stage of optical qubits is an essential task in all the experimental setups employed for the test and demonstration of Quantum Optics principles. We consider a deterministic protocol for the preparation of qubits as a superposition of vacuum and one photon number states, which has the advantage to reduce the amount of resources required via phase-sensitive measurements using a local oscillator (`dyne detection'). We investigate the performances of the protocol using different phase measurement schemes: homodyne, heterodyne, and adaptive dyne detection (involving a feedback loop). First, we define a suitable figure of merit for the prepared state and we obtain an analytical expression for that in terms of the phase measurement considered. Further, we study limitations that the phase measurement can exhibit, such as delay or limited resources in the feedback strategy. Finally, we evaluate the figure of merit of the protocol for different mode-shapes handily available in an experimental setup. We show that even in the presence of such limitations simple feedback algorithms can perform surprisingly well, outperforming the protocols when simple homodyne or heterodyne schemes are employed.
%
\end{abstract}

\maketitle


\section{Introduction
\label{sec:intro}
}

Linear Optics Quantum Computation (LOQC) has proved to be an effective platform for both demonstrations of Quantum Mechanics principles \cite{Zeilinger1999} and the development of applications in the fields of communications \cite{Holevo2012}, quantum key distribution \cite{Gisin2007}, and metrology \cite{Giovannetti2011}. It involves the preparation, propagation and measurements of optical quantum bits (qubits) through a network of linear optical elements. Because LOQC is inherently non-deterministic, the consumption of quantum resources is a critical issue in such systems.

The conditions the network design and resource consumption are strongly guided by the qubit encoding scheme \cite{Ralph2005}. 
These can be classified into two main categories --- \emph{single rail logic} or \emph{dual rail logic}. In the former, a single-mode system is used, and the qubit is encoded as a superposition of the single mode $n$-photon Fock states $\ket{0_L}=\ket{0}$ and $\ket{1_L}=\ket{1}$. In the latter, a two-mode system is used, and the qubit is encoded as superposition of the logic mode $\ket{0_L}=\ket{0}\ket{1}$ and $\ket{1_L}=\ket{1}\ket{0}$, with $\ket{n}\ket{m}$ the tensor product between two single-mode Fock states. 

To date, many important experimental demonstrations of optical quantum information processing concepts have been made with dual-rail systems, using for example polarisation modes \cite{OBrien2003,Pittman2002}, spatial modes \cite{Politi2008} or "time-bin" modes \cite{Humphreys2013}.  Single-rail implementations of LOQC are of fundamental interest because of the different, and potentially less resource-intensive, ways in which errors occur and are corrected \cite{Lund2002}.  However, single-rail LOQC is yet to be actively explored in an experimental setting because of the relative experimental difficulty associated with producing single-rail superposition states compared to dual rail schemes.  

The original proposals \cite{Lund2002,Pegg1998}  and pioneering demonstration-of-principle experiments \cite{Babichev2003} to prepare single-rail qubits considered \emph{non-deterministic} single qubit gates. But, by virtue of being non-deterministic, those protocols required many resources and had low probability of success.  Instead, it was shown \cite{Ralph2005} that adding dyne detection and feedback makes it possible to prepare an arbitrary single-rail qubit via a \emph{deterministic} protocol, thereby reducing dramatically the resource consumption and making single-rail LOQC an important quantum information architecture to explore.

In this paper, we study that deterministic protocol for the preparation of a qubit in single rail encoding in more detail and from an experimental perspective. The protocol leverages a canonical phase measurement, which in principle could be implemented with an adaptive homodyne detection \cite{Wiseman1996}, but in an experimental setup cannot be perfectly implemented due to physical limitations and non-idealities. 

We first define a figure of merit for the protocol performance in producing the qubit, and we evaluate it when the homodyne, heterodyne of adaptive homodyne phase measurements are employed. As expected, adaptive homodyne measurement allows  perfect preparation of the required qubit \cite{Ralph2005}. We then consider non-idealities and limitations that the phase measurement can exhibit, such as delay or limited resources in the feedback strategy. We provide an analytical formulation for the figure of merit, and evaluate it both for constant gain feedback, optimized for any given delay, and for 
a simple time-varying gain (piecewise constant with two values), again optimized for any given delay. Evaluating the performances for the different optical mode-shapes that are handily available in an experimental setup, we show that simple feedback algorithms can perform surprisingly well, even in the presence of time-delays in the feedback loop.   

The paper is arranged as follows. We define fidelity as the figure of merit for phase measurement in Section \ref{sec:protocol} and then show how to calculate that figure of merit for dyne measurements in Section \ref{sec:phase}.  We then present an analytically calculable approximation to the fidelity in Section \ref{sec:expression}. Based on those results, we evaluate in Section \ref{Sec:Evalopt} the fidelity for a variety of experimentally-motivated feedback algorithms when they are applied to a range of different mode-shapes for the input fields.  We conclude in Section \ref{sec:conc} and present supplementary mathematics in the Appendices.

\section{Single-rail qubit preparation} 
\label{sec:protocol}

\subsection{Ideal phase measurement}

\begin{figure}
\centering
\includegraphics[width=\columnwidth]{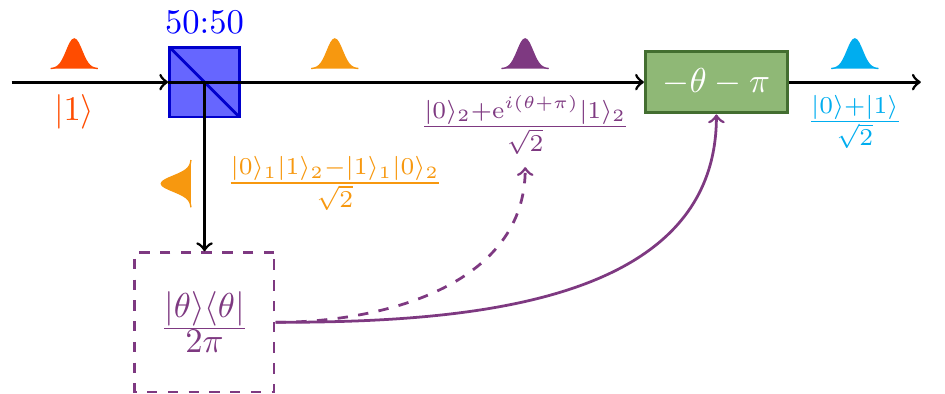}
\caption{(Color online) Scheme of the deterministic protocol for the preparation of qubit \eqref{equal}. A single photon (red) is sent through a 50:50 beam splitter. A canonical phase measurement (dashed block) is performed on one output port, making the entangled state (orange) to collapse on the other port. The phase of this collapsed state is corrected with a phase modulator (green), in order to obtain the desired qubit (cyan).}
\label{fig:scheme5050}
\end{figure}

To begin, we recapitulate the protocol presented in \cite{Ralph2005} for the 
deterministic preparation of an arbitrary pure qubit state 
within the single rail photonic encoding (see Fig. \ref{fig:scheme5050}). 
Consider for the moment the preparation of the equal superposition state, i.e.
\beq
\ket{\qubit} = \frac{\ket{0} + \ket{1}}{\sqrt{2}}.
\label{equal}
\eeq 
Assuming we can generate a single photon deterministically, we can send this 
through a 50:50 beam splitter, generating two out-going entangled modes in the state
\beq
\ket{\Psi}=\frac{\ket{0}_1\ket{1}_2 - \ket{1}_1\ket{0}_2}{\sqrt{2}},
\eeq 
where the subscripts $1,2$ indicate the output ports of the beam splitter.

Suppose that on the first mode we perform a canonical phase measurement described by the POM
\beq
\hat{E}_{\rm can}(\theta) = \frac{\pure{\theta}}{2 \pi},\ \ket{\theta} = \ket{0}_1 + e^{i\theta} \ket{1}_1,
\label{canonicalPhaseMeasurement}
\eeq 
which is normalized according to 
\beq
\int_0^{2\pi} d\theta \ \hat{E}_{\rm can}(\theta) = \pure{0} + \pure{1} = \hat 1
\eeq
where the identity $\hat 1$ is restricted to the subspace of interest, spanned by $\{\ket{0}_1, \ket{1}_1\}$.	 
As a result of the measurement, we obtain the outcome $\theta$, drawn randomly with a uniform probability distribution in the range $[0, 2\pi)$. After this measurement, the quantum state on the other mode 
collapses to 
\beq
\ket{\psi}_2 = \frac{ \ket{1}_2 -e^{-i \theta}\ket{0}_2 }{\sqrt{2}} = \frac{ \ket{0}_2 + e^{i (\theta+\pi)}\ket{1}_2 }{\sqrt{2}}.
\label{collapsed}
\eeq 
Then, by using classical feed-forward from the phase measurement, 
we can undo the random phase $\theta+\pi$ by passing the state through a phase modulator 
to create \eqref{equal}. 

This protocol is easily generalizable to arbitrary qubit states. We can prepare non-equal superpositions by employing
a beam splitter with intensity reflectivity $\eta$, to get at the output of the beam splitter the quantum state
\beq
\frac{ \sqrt{\eta}\ket{0}_1\ket{1}_2 - \sqrt{1-\eta}\ket{1}_1\ket{0}_2}{\sqrt{2}}.
\eeq 
Again, measuring the first mode with a canonical phase measurement, we can correct the phase on the second mode, 
and apply any desired additional phase $\varphi$, to obtain
\beq
\ket{\psi} = \sqrt{1-\eta} \ket{0}_2 + e^{i \varphi} \sqrt{\eta} \ket{1}_2.
\label{genericQubit}
\eeq

\subsection{Non-ideal phase measurement}

It is not easy in optics to implement the canonical POM \eqref{canonicalPhaseMeasurement}, 
as we will investigate. The problem of efficiency is an obvious one, but this is simply
equivalent to loss, and so can be incorporated into the final analysis by applying loss to the state. 
A more interesting problem is how to project onto the state $\ket{\theta}$, with its equal superposition 
of $\ket{0}$ and $\ket{1}$. This equal superposition is essential for ensuring that all outcomes 
lead to a state that is only a phase shift away from the desired qubit state. More generally, 
an optical measurement will involve projecting onto an unnormalized state
\beq
\ket{\R} = \ket{0} + \R \ket{1},
\eeq 
where $\R \in \mathbb{C}$. The POM describing such a measurement is 
\beq
\hat{E}(\R) = \wp_{\rm ost}(\R) \pure{\R}, \label{pom}
\eeq
where $\wp_{\rm ost}(\R)$ is a positive, normalized distribution satisfying 
\beq \label{post1}  
\int \wp_{\rm ost}(\R) \ \d^2 \R = \int \wp_{\rm ost}(\R) \ |\R|^2 \ \d^2 \R = 1.
\eeq 
This ensures that the POM obeys the completeness relation
\beq
\int \hat{E}(\R) \ \d^2 \R = \hat{1}
\label{completenessRelation}
\eeq
as required. 

The notation  $\wp_{\rm ost}(\R)$ is because we call this an {\em ostensible} 
probability distribution for $R$ \cite{Wiseman1996}. 
The {\em actual} probability distribution for $R$ is, for an input state $\ket{\Psi}$,  
\beq
\Prob{\R}  = \tr[]{\pure{\Psi} \hat{E}(\R)} = \wp_{\rm ost}(\R) \braket{\Psi}{\R}\braket{\R}{\Psi}.
\label{BornRule}
\eeq 
As we will see, the properties of $\wp_{\rm ost}$ are critical to the performance of the protocol. 
Given the result $\R$, the best estimate of the phase is always 
\beq
\theta = \arg \R.
\label{estTheta}
\eeq
Thus, the POM for this phase measurement is 
\begin{align}
\hat{E}(\theta) d\theta &= d\theta \int_0^\infty 	d|\R| \ \hat{E}(\R) \nonumber\\
&=  \frac{d\theta}{2\pi}\sq{\pure{0} + \merit \ro{\ket{1}\bra{0}e^{i\theta} + \ket{0}\bra{1}e^{-i\theta} } + \pure{1}} \nonumber \\
&= \merit\times \hat E_{\rm can}(\theta) \ d\theta  + (1-\merit) \times \hat{1}\ \frac{d\theta}{2\pi},
\end{align}
where 
\beq \label{defFM}
\merit=\ave{|\R|}_{\ost} \equiv \int d^2 \R \ |\R| \ \wp_{\rm ost}(\R)
\eeq 

Clearly, $\merit$ is a figure of merit expressing how close the measurement is to canonical.  
It is also directly related to the fidelity of the qubit preparation protocol above. 
Considering the 50:50 case for simplicity, and defining $\mathcal{U}_2(\alpha)\bullet = e^{i\alpha \hat a\dg_2 \hat a_2} \bullet e^{-i\alpha \hat a\dg_2 \hat a_2}$, 
the final quantum state is
\begin{align}
\rho_2 & = \int \mathcal{U}_2(-\arg\R-\pi) \ \tr[1]{\pure{\Psi} \hat{E}(\R)}  \d^2\R \label{eq0}\\
& = \frac{\pure{0}_2 + \merit\ro{\ket{0}\bra{1}_2 +   \ket{1}\bra{0}_2 } + \pure{1}_2 }{2} \label{eq2}.
\end{align} 
This has a purity of   
$\tr{(\rho_2)^2} = \rnfrac{1+F^2 }{2}$ and a fidelity with the desired state of 
$\bra{\psi}\rho_2\ket{\psi} = \rnfrac{1 + F}{2}$.  Thus we can use \eqref{defFM} as figure of merit to evaluate the protocol.
 
\section{\label{sec:phase}Phase measurement by dyne detection}

\begin{figure}
\centering
\includegraphics{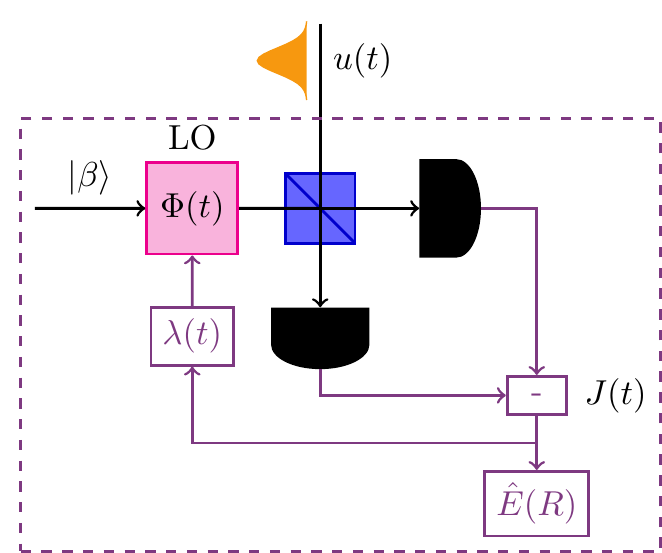}
\caption{\label{fig:phaseMeasurement}(Color online) Scheme of a dyne phase measurement (homodyne, heterodyne or adaptive), corresponding to the dashed block of Fig. \ref{fig:scheme5050}. The incoming beam (orange) is mixed with a strong local oscillator (pink block) with phase $\X(t)$, and the quadratures are measured. The difference of the photocurrents $\J(t)$ is post-processed to obtain the outcome $\R$. In the case of adaptive phase measurement, the phase of the local oscillator is adjusted during the measurement with the gain $\y(t)$.}
\end{figure}

The non-ideal phase measurement for single rail qubits described in the preceding section 
can be applied to all forms of ``\emph{dyne}'' \cite{WisKil98, Wiseman2009} detection schemes: homodyne, heterodyne, or adaptive dyne detection. Following \cite{Wiseman2009}, we can describe all of them in the same framework of the quantum trajectories (see Fig. \ref{fig:phaseMeasurement}).

Consider a single optical mode with mode-shape $\u(t)$, $\u(t) \geq 0$ (shortened later as $\ut$), 
for $t \in {\mathbb R}$, such that 
\beq
\U(t) = \int_{-\infty}^t \u(s) \ds, \quad \U(\infty)=1.
\eeq
Coming out from the first port of the beam splitter, the optical mode  $\ket{\psi}_1$ is mixed through a balanced beam splitter with a mode-matched local oscillator (LO) with phase $\X(t)$ (shortened to $\Xt$). In the limit of a large LO amplitude, all of the information is contained in the 
difference $\J(t)$ between the photocurrents detected at each output port of the beam splitter. 
We will normalize this so that for a vacuum input ($\ket{\psi}_1=\ket{0}$),  
\beq \label{ostJ}
\J(t) \dt = \dW(t) ,
\eeq
where this denotes a Wiener increment satisfying
\begin{align}
&\langle \dW(t)\rangle = 0, \\
&\dW(s) \dW(s) = \dt.
\end{align}

Not only does Eq.~\ref{ostJ} describe the photocurrent statistics 
for a vacuum input, it  also defines the {\em ostensible} statistics for an arbitrary qubit state. 
That is, it is the appropriate equation to use to define the 
{\em ostensible} distribution $\wp_{\rm ost}(\R)$ introduced above. 
As we have seen, the properties of $\wp_{\rm ost}(\R)$ are all that are needed 
to define the quality of the preparation. The relation between $\J(t)$ and $\R$ is \cite{Ralph2005,WisKil98,WisKil97}
 \beq
 \R = \R(\infty)\;;\;\; \R(t) = \int_{-\infty}^t e^{i \X(s)} \sqrt{\u(s)} \J(s) \ds. \label{defAt0}
 \eeq
Note that the LO phase $\X(t)$ is completely arbitrary, and may even be
 made dependent upon the measurement record up to time $t$,
$\mathbf{\J}_t=\{\J(s): s\leq t\}$.  This makes the measurement {\em adaptive} and allows for better phase estimation than is otherwise possible 
\cite{WisKil98,WisKil97}.

In the following, we use the notation $\E(t)$ instead of $\J(t)$, for calculating ostensible 
probabilities, with $\E(t) \dt$ equal to the Wiener increment $\dW(t)$. 
Then the figure of merit is 
\beq
\merit = \ave{ \st{\int_{-\infty}^{\infty} {e}^{i \X(s)} \sqrt{\u(s)} \E(s) \ds }},
\eeq
where for adaptive detection $\X(t)$ will depend upon 
${\boldsymbol \E}_t~=~\{\E(s):s\leq t\}$. 

\subsection{Homodyne measurement}

In the homodyne detection, the LO phase $\X(t)$ is constant in time, i.e. $\X(t) = \X_0$. 
In this case, the definition of $\R$  becomes
\begin{align}
\R &= e^{i \X_0} X, \label{homoR}
\end{align}
with 
\beq
X = \int_{-\infty}^\infty  \sqrt{\u(s)} \E(s) \ds.
\eeq \blk
This is a real Gaussian random variable with zero mean and hence, 
as required by \erf{post1}, unit variance.
Therefore, the ostensible distribution for $X$ is 
\beq
\ProbOst{X} \d X = \frac{1}{\sqrt{2 \pi}} e^{-\xfrac{X^2}{2}} \d X ,
\label{probX}
\eeq
and we can calculate 
\beq
\merit = \int_{-\infty}^{+\infty} |X| \ProbOst{X} \d X = \sqrt{\frac{2}{\pi}} \approx 0.797
\label{exactX}
\eeq

\subsection{Heterodyne measurement}
In  heterodyne detection, the LO is detuned, i.e. outside the bandwidth of the optical mode. That is, 
the phase $\X(t)$ is linearly increasing in time at a rate $\Delta~\gg~{\rm max}_t [u(t)]$.  
By definition \eqref{defAt0} we get
\begin{align}
\R &=  \int_{-\infty}^{\infty}  e^{i \X_0 + i\Delta t} \sqrt{\u(s)} \E(s) \ds.
\end{align}
This is a complex Gaussian random variable with zero mean and 
in the limit of $\Delta \to \infty$ gives a rotationally symmetric  
variable $A$ with 
\beq
\ProbOst{A} \d^2 A = \frac{1}{\pi} e^{-|A|^2} \d^2 A.
\label{probA}
\eeq 
From this, we can evaluate 
\beq
\merit = \int |A| \ProbOst{A} \d^2 A = \frac{\sqrt{\pi}}{2} \approx 0.886,
\label{exactA}
\eeq 
showing that heterodyne results in a better phase measurement than homodyne.

\subsection{Adaptive measurement}

To do even better than heterodyne detection it is necessary to implement 
 adaptive dyne detection. In particular, Ref.~\cite{Wiseman1995} introduced the 
 following adaptive algorithm:  
\beq
\X(t) = \arg \R_t + \frac{\pi}{2}.
\label{algorithm}
\eeq 
This gives the stochastic differential equation for $\R_t$
\beq
\d \R_t= i \frac{\R_t}{|\R_t|} \sqrt{\u(t)} \J(t) \dt
\eeq
and the deterministic differential equation for $|\R_t|^2$
\beq
\d |R_t|^2 = u(t) dt.
\eeq 
Solving this, we get $\R(t) = e^{i\theta(t)}$, with 
\beq \label{thetat}
\theta(t) =   \int_{-\infty}^{\infty}\sqrt{\frac{\u(s)}{\U(s)}} \  \J(s) \ds. 
\eeq
The ostensible statistics for $\theta = \theta(\infty)$ are those of 
\beq \label{ostthetat}
\theta  =   \int_{-\infty}^{\infty}\sqrt{\frac{\u(s)}{\U(s)}} \  \E(s) \ds, 
\eeq
which has a divergent variance because $\U({-\infty})=0$. 
This means that, as in the heterodyne case, $\wp_{\rm ost}(\R)$ is rotationally 
invariant. But in this case, since $|R|=1$, we have immediately 
\beq
\merit = \ave{|R|}_{\rm ost} = 1,
\eeq
and hence this adaptive dyne protocol realizes the canonical phase measurement. \blk

\subsection{Non-optimal adaptive measurement}

The adaptive algorithm \eqref{algorithm} gives the optimal expression \eqref{adaptiveGain} to use in the feedback loop for  our qubit preparation protocol. Remembering that \erf{thetat} is equal to $\arg(R_t)$, 
\erf{algorithm} can be re-expressed as integral feedback
\beq \label{optgain}
\X(t) =  \frac{\pi}{2} + \int_{-\infty}^{t} \y_{\rm opt}(s) \J(s) \ds   ,
\eeq
with the time-dependent gain  
\beq
\y_{\rm opt}(t) = \sqrt{\frac{\u(t)}{\U(t)}}.
\label{adaptiveGain}
\eeq

Experimentally, not only it is difficult to implement a time-dependent gain in the feedback loop, but it is impossible to implement a  completely divergent gain (as required for most mode-shapes when the pulse first turns on). 
Also, \erf{algorithm} assumes 
zero time delay in the feedback loop, which is unrealistic. These considerations 
 motivates the focus of the remainder of this paper, which is to consider 
 a feedback protocol of the form 
\beq
\X(t) = \frac{\pi}{2} + \int_{-\infty}^{t-\delay} \y(s) \J(s) \ds,
\label{defXt}
\eeq 
that is, a  LO phase which depends on the record measurement only up to $t-\delay$ with a gain function $\y$ subject to some physical restrictions. 

In the absence of delay $\delay$, applying the feedback \eqref{algorithm},
 we can attain $\merit=1$. In presence of the delay the optimal gain $\y(t)$ to apply for the feedback 
is not known. Indeed, it is likely that the optimal feedback in this case cannot be written in the form \eqref{defXt}, 
 and instead some more complex algorithm would be optimal. However, the expression \eqref{defXt} 
 still allows for a great deal of flexibility, even when choosing 
 $\y(s)$ from a restricted class of functions, as we will find in Sec.~\ref{Sec:Evalopt}.

With such a general expression as \erf{defXt} there will no longer be any simple expressions for the ostensible 
statistics of $\R_t$. Instead, one has to evaluate the moments of the stochastic integral  
\beq \label{defA}
\R = \int_{-\infty}^{\infty} e^{i \X(s)} \sqrt{\u(s)} \E(s) \ds.
\eeq
with $\X(t)$ given by  
\beq
\X(t) = \frac{\pi}{2} + \int_{-\infty}^{t-\tau} \y(s) \E(s) \ds.
\label{ostXt}
\eeq
We will see in the next Section how to develop an approximation for $\merit$ from 
the ostensible moments of $\R$. 

\section{Approximate Figure of Merit 
\label{sec:expression}
}
In this section, we develop an approximation, $\am$, for the figure of merit $\merit$. 
Due to the non-analytic nature of $|\R|$ as function of $\R$, we cannot calculate exactly $\merit$ even given an expression for $\R$. However, since we are interested in the limit of good phase measurements, 
$\ave{|\R|}_{\ost} \approx 1$, we can approximate $\R$ to its second order in a Taylor series 
about $\R = 1$ to get \blk 
\begin{align}
\ave{|\R|}_{\ost} & = \ave{\sqrt{|1+(|\R|^2-1)|}}_{\ost} \\
& \approx 1 + \frac{\ave{|\R|^2-1}}{2}-\frac{\ave{(|\R|^2-1)^2}_{\ost}}{8}\\
& = 1 - \frac{\ave{|\R|^4}_{\ost}-1}{8}
\label{approximation}
\end{align}
where we have used \erf{post1}. Thus we define our new figure of merit 
\beq \label{defam}
\am =  \frac{9 - \ave{|\R|^4}}{8}. 
\eeq 
Here, and in the remainder of this paper, we have dropped the `ost' subscript 
on the average, since all averages we will be considering are ostensible averages. 

Before determining $\am$ for non-ideal adaptive dyne measurement, we first  
apply the approximation \eqref{approximation} to the cases where we have already evaluated 
$\merit = \ave{|\R|}$, in order to test the validity of the approximation. 
In the case of homodyne detection, we have, from \erf{probX}, $\ave{|X|^4}=3$, and hence 
\beq
\ave{|\R|} \approx \am = 1 - \frac{\ave{|X|^4}-1}{8} = 0.75
\label{approxX}
\eeq 
For the heterodyne measurement, we obtain $\ave{|A|^4}=2$ from \erf{probA}, 
and hence
\beq
\ave{|\R|} \approx \am = 1 - \frac{\ave{|A|^4}-1}{8}= 0.875
\label{approxA}
\eeq 
Finally,   in the case of ideal adaptive dyne detection we get $|\R|=1$ deterministically, and hence we get exactly 
\beq
\ave{|\R|} = \am = 1.
\eeq 
Comparing the exact values \eqref{exactX} and \eqref{exactA} with the approximations \eqref{approxX} and \eqref{approxA}, we find good agreement, as shown in Table \ref{exactApproxTable}. As we can see, the adaptive homodyne outperforms the other schemes, and therefore constitutes a starting point for the implementation of a phase measurement in the presence of experimental limitations. 
In the remainder of the paper we use $\am$, rather than $\merit$, to compare the different adaptive schemes. 

\begin{table}
\centering
	\begin{tabular}{ccc}
	Measurement & Exact & Approx.\\
	\hline
	Homodyne & $\displaystyle \sqrt{\xfrac{2}{\pi}} \approx 0.797 $  & 0.75 \\
	Heterodyne & $\displaystyle \sqrt{\xfrac{\pi}{4}} \approx 0.886 $ & 0.875 \\
	Adaptive Homodyne & 1 & 1 \\
	\end{tabular}
	\caption{Exact and approximated performances for different dyne measurements.}
	\label{exactApproxTable}
\end{table}

\section{Evaluating and optimizing $\am$} \label{Sec:Evalopt}

In this section we evaluate and maximize the approximate figure of merit $\am$ when the phase measurement is implement with an adaptive algorithm which exhibits limitations such as delay or restrictions in the resources of the feedback scheme. These non-idealities reflect the problems that we face in an experimental setup.

We first show the general, analytical expression for $\am$ as functional of the gain $\y(t)$ and the pulse shape $\u(t)$. The technical details of the derivation are included in \ref{sec:fidelDyne}, here in Sec.~\ref{sec:analytical} we report only the final expression. 
We then evaluate $\am$ as a function of $\delay$, for four different mode-shapes (see Sec.~\ref{sec:mode-shapes}), 
with two different types of gain. The simplest strategy for the feedback loop is constant gain, which we consider 
in Sec.~\ref{sec:constantGain}. As one step (and three parameters) beyond constant gain 
we then consider piecewise-constant gain, with a single discontinuity, in Sec.~\ref{sec:piecewise}. 
While this feature (a discontinuity in the gain) is not particularly realistic, we expect the results 
for that section to give a reasonable idea of how much improvement may be gained by 
the most basic improvements over the simplest case of constant gain. 

\subsection{Analytical expression for $\am$} \label{sec:analytical}

We first consider the case of zero delay in the feedback loop, i.e. $\tau=0$. The final expression that we obtain reads
\begin{align}
\am & =  1- \int_{-\infty}^{+\infty} \dt \int_{-\infty}^{t_{-}} \ds \ \ut \us \rnfrac{1+{e}^{\textstyle -2\int_s^t \ytau^2 \ \d \tau}}{4}  \nonumber \\
& \qquad +  \int_{-\infty}^{+\infty} \dt \int_{-\infty}^{t_{-}} \ds \int_{-\infty}^{s_{-}} \dv \ \ut \sqrt{\us \uv}\ys \yv {e}^{\textstyle -2\int_s^t \ytau^2 \ \d \tau} {e}^{\textstyle -\int_v^s \ytau^2 \ \d \tau/2}. 
\label{final2}
\end{align} 
If we substitute the optimal adaptive gain $\y_{\rm opt}(t)$ of \eqref{adaptiveGain}, the two integral terms in expression \eqref{final2} cancel identically, giving $\am=1$ as it should be. Given how complicated \eqref{final2} is, this is a powerful check on its correctness.

We then consider the case with delay, $\tau > 0$. The expression in this case becomes even more complicated, that is
\begin{align}
\am & =1 -    \int_{-\infty}^{+\infty} \dt \int_{0}^{+\infty} \dD \ \ut \utD \rnfrac{1+{e}^{-2\int_{t-\Delta-\delay}^{t-\delay} \ys^2 \ds}}{4}   \notag  \\
&  + {2} \int_{-\infty}^{+\infty} \dt \int_\delay^{+\infty} \dD \int_0^\delay \dd \ \ytD \ytDd \ut \sqrt{\utD \utDd} {e}^{-2\int_{t-\Delta-\delay}^{t-\delay} \ys^2 \ds}  {e}^{-\frac{1}{2}\int_{t-\Delta-\delta-\delay}^{t-\Delta-\delay} \ys^2 \ds}  \label{final3}   \\
& + \int_{-\infty}^{+\infty} \dt \int_\delay^{+\infty} \dD \int_\delay^{+\infty} \dd \ \ytD \ytDd \ut \sqrt{\utD \utDd} {e}^{-2\int_{t-\Delta-\delay}^{t-\delay} \ys^2 \ds}  {e}^{-\frac{1}{2}\int_{t-\Delta-\delta-\delay}^{t-\Delta-\delay} \ys^2 \ds}   \notag
\end{align}
As a check, the limit for $\delay \to 0$ returns the expression \eqref{final2}. 
 
\subsection{Modeshapes} \label{sec:mode-shapes}

In order to test the performances of the protocol against the delay, we consider four different mode-shapes: 
\begin{itemize}
\item {rectangular shape}
\beq
\u(t) =\begin{cases} 
1, & \mbox{if } 0<t<1 \\
0, & \mbox{otherwise}
\end{cases};
\label{rect}
\eeq

\item {bilateral exponential shape}
\beq
\u(t)=\frac{\kappa}{2}e^{-\kappa|t|};
\label{bil}
\eeq

\item {falling unilateral exponential shape} 
\beq
\u(t) =\begin{cases} 
ke^{-k t}, & \mbox{if } t\geq0 \\
0, & \mbox{otherwise}
\end{cases};
\label{expP}
\eeq

\item {rising unilateral exponential shape} 
\beq
\u(t) =\begin{cases} 
ke^{k t}, & \mbox{if } t\leq0 \\
0, & \mbox{otherwise}
\end{cases}.
\label{expN}
\eeq
\end{itemize}

Some of these mode-shapes arise naturally in an experimental setup.  The rectangular mode-shapes can easily be obtained to a very good approximation with 
weak coherent light, which could be used to test phase measurements at a single-photon level.  The bilateral mode-shape appears with a single signal photon coming from spontaneous parametric down conversion (SPDC), conditioned on a click in the idler mode at time $t=0$. The falling unilateral exponential shape is the mode of a photon from an initially excited atom.  It is less obvious how the rising unilateral exponential shape might be obtained experimentally, but has a very nice theoretical property as will be explained below. 

\blk When we introduce a delay $\delay$, we want to compare what effect it has on schemes
using these four different mode-shapes. To make a fair comparison we need the characteristic 
times of these pulses to be the same. Since all but one of the above mode-shapes do not have 
compact support, we cannot simply use the pulse duration. Instead, we adopt the following 
measure of characteristic duration: 
\beq
\w = \sq{\int_{-\infty}^{+\infty} \u^2(t) \dt}^{-1}
\label{w}
\eeq
For the four mode-shapes above, this evaluates respectively to $1$, $\frac{4}{\kappa}$, $\frac{2}{k}$, and $\frac{2}{k}$. Thus we normalize these all to have $\w=1$ by setting $\kappa=4$ and $ k= 2$.

\subsection{Constant gain} \label{sec:constantGain}

We first consider the case $\y(s)=\y$. 
We substitute the various mode-shapes into  \erf{final3}, in order to obtain  analytical expressions 
for $\am$ as a function of the delay $\delay$ and the gain $\y$.  These expressions (which are lengthy and not very informative) 
are reported in \ref{AppendixB}. We then numerically optimize over $\y$.  The 
results are shown in Fig.~\ref{fig:constantGainPerformances}, along with 
the performance of  heterodyne detection (the best non-adaptive scheme). 
 
\begin{figure}
\centering
    \centering
    \subfloat[a][\label{fig:constantGainPerformances}Constant gain]{
            \includegraphics[width=0.49\textwidth]{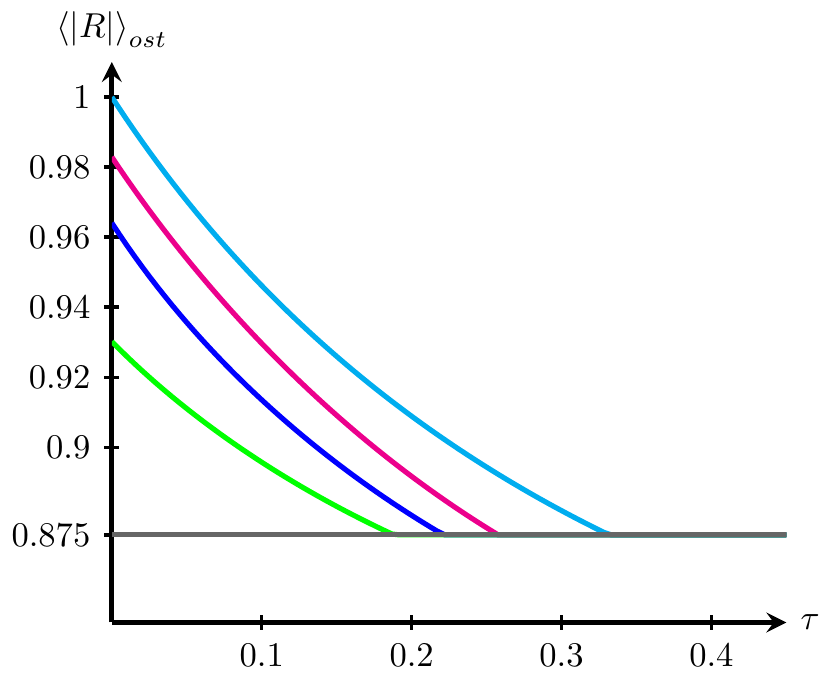}
    }
    \subfloat[b][\label{fig:twoGainsPerformances}Piecewise constant gain]{
            \includegraphics[width=0.49\textwidth]{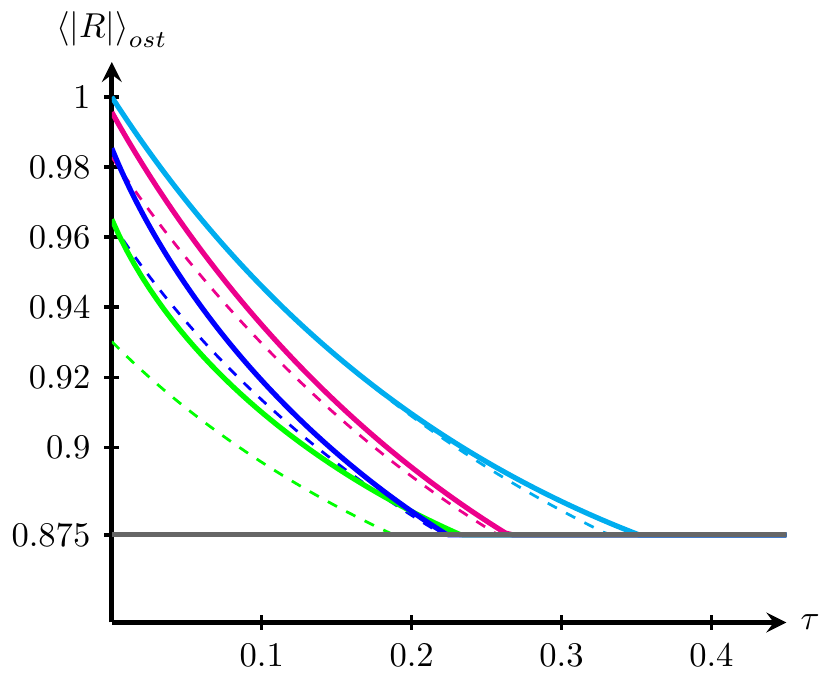}
    }
\caption{\label{fig:constantGainPerformances}(Color online) Performances of constant gain (a) and piecewise constant gain (b) adaptive scheme with respect to time-delay $\delay$, for different mode-shape $\u(t)$. The curves, from top to bottom, are rising unilateral exponential (cyan), bilateral exponential (magenta), rectangular (blue), falling unilateral exponential (green) and heterodyne (gray), plotted for comparison. Dashed lines in (b) report the corresponding solid lines in (a).}
\end{figure}

For all the mode-shapes, adaptive measurement out-performs heterodyne detection at zero delay, 
but the performances decrease monotonically with delay, as expected. The rising exponential mode-shape \eqref{expN} outperforms all other modes. In fact, for $\delay=0$, this mode-shape 
yields a perfect phase measurement for constant gain. This can be verified analytically, as 
 the optimal solution of \eqref{adaptiveGain} is $\lambda = 4\sqrt{2}/\kappa$ for this mode-shape. 
The bilateral exponential is second-best, which is heartening for experiments based upon 
SPDC. The order in terms of $\am$ at $\delay=0$ is preserved for $\delay > 0$,  
and the  rising exponential mode-shape is particularly robust to delay, 
with constant gain feedback still showing an improvement over 
heterodyne detection for $\delay$ as large as $1/3$ of the characteristic pulse time.

\begin{figure}
\centering
\includegraphics{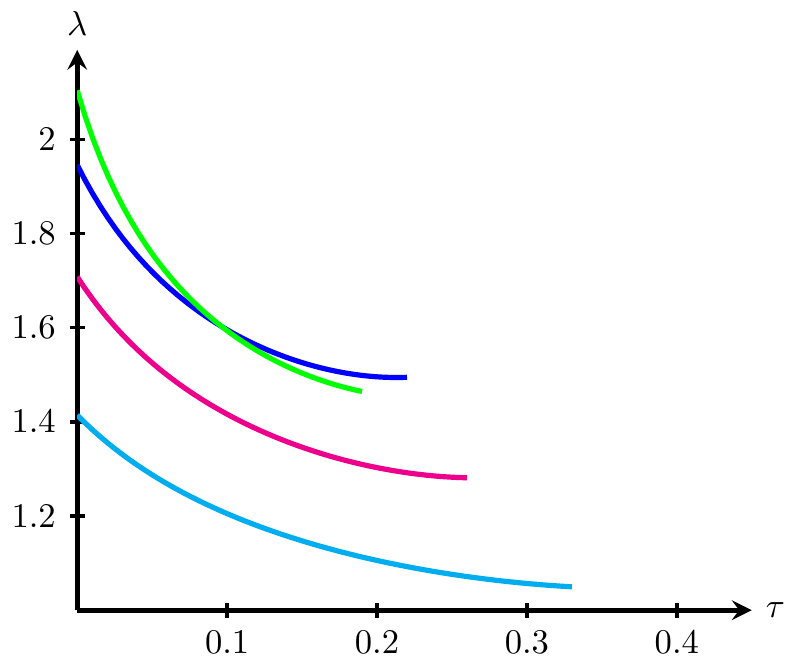}
\caption{\label{fig:constantGain}(Color online) Plot of the optimized constant gain for the feedback with respect to time-delay $\delay$, for different mode-shape $\u(t)$. The curves, from top to bottom at $\tau=0$, are falling unilateral exponential (green), rectangular (blue), bilateral exponential (magenta) and rising unilateral exponential (cyan).}
\end{figure}

In Fig. \ref{fig:constantGain}, the values of constant gain employed in the feedback are plotted with respect to
the delay, for the different mode-shapes. A common trend can be identified: as the delay increases the optimal $\y$ decreases. Also, the size of the optimal gain is, for the most part, inversely related to the effectiveness $\am$ of the phase measurement. 

\subsection{Piecewise constant gain} \label{sec:piecewise}

The results above were obtained under the assumption of constant gain. This is the simplest model to consider, 
having only a single parameter to optimize over. It is natural to ask how much improvement is possible with a 
slightly more complicated model, as this will give a guide to what might be expected with even more careful tailoring of the feedback algorithm. This motivates considering a piecewise constant gain $\y(t)$, 
with three parameters:
\beq
\y(t) = \begin{cases}
\y_1 & \textrm{for }t \leq \tl \\
\y_2 & \textrm{for }t > \tl
\end{cases}
\label{twoGains}
\eeq
Again, we find the analytical expression for $\am$ for the four different mode-shapes by substituting the feedback \eqref{twoGains} into \erf{final2}. The analytical expression for $\am$ are not reported here due to their length and complexity. Next, we optimize over the values $\y_1,\ \y_2$ and $\tl$.   

In Fig. \ref{fig:twoGainsPerformances} the performances of the phase measurement with the optimized feedback are plotted in continuous lines, while in dashed lines the previous performances with constant gains are reported. 
We can see quite substantial improvements for the bilateral, rectangular and falling unilateral exponential mode-shapes, although the improvement lessens for longer delays. The opposite is true for the rising exponential mode-shape, as expected since constant gain is optimal at $\delay=0$ for this case.

\begin{figure*}
    \centering
    \subfloat[a][\label{fig:twoGainsR}Rectangular shape]{
            \includegraphics[width=0.49\textwidth]{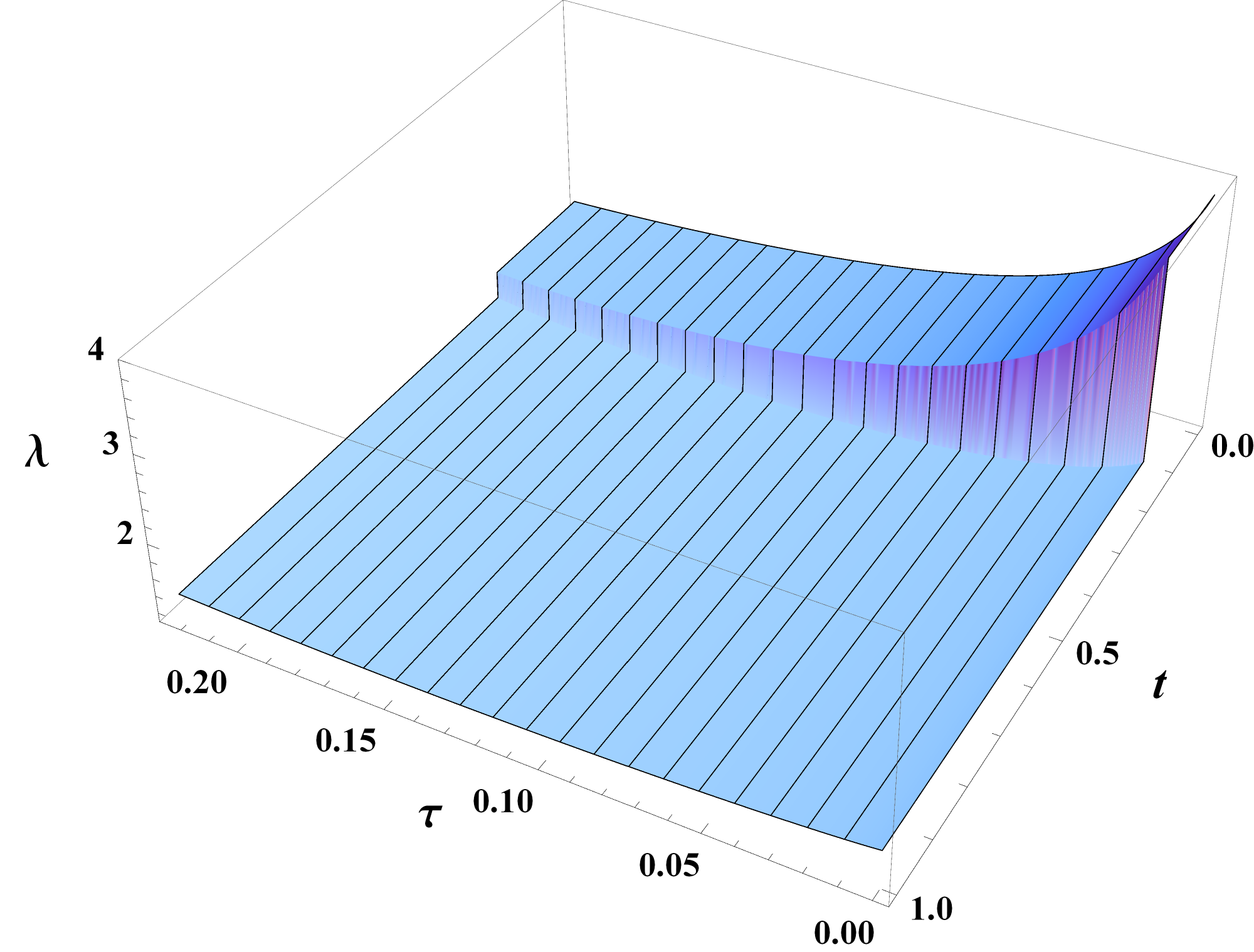}
    }
    \subfloat[b][\label{fig:twoGainsB}Bilateral exponential shape]{
            \includegraphics[width=0.49\textwidth]{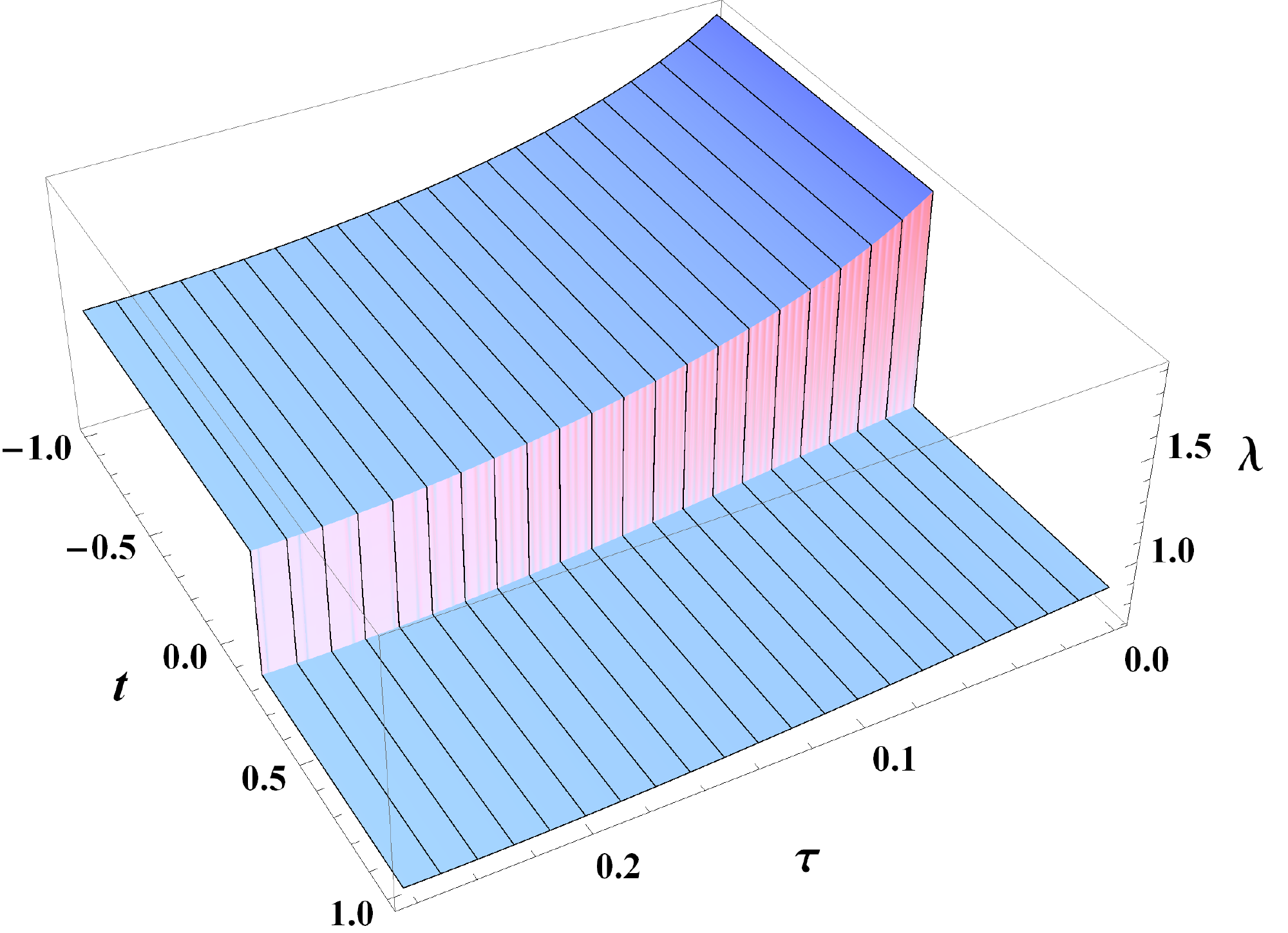}
    }

    \subfloat[c][\label{fig:twoGainsEP}Falling exponential shape]{
            \includegraphics[width=0.49\textwidth]{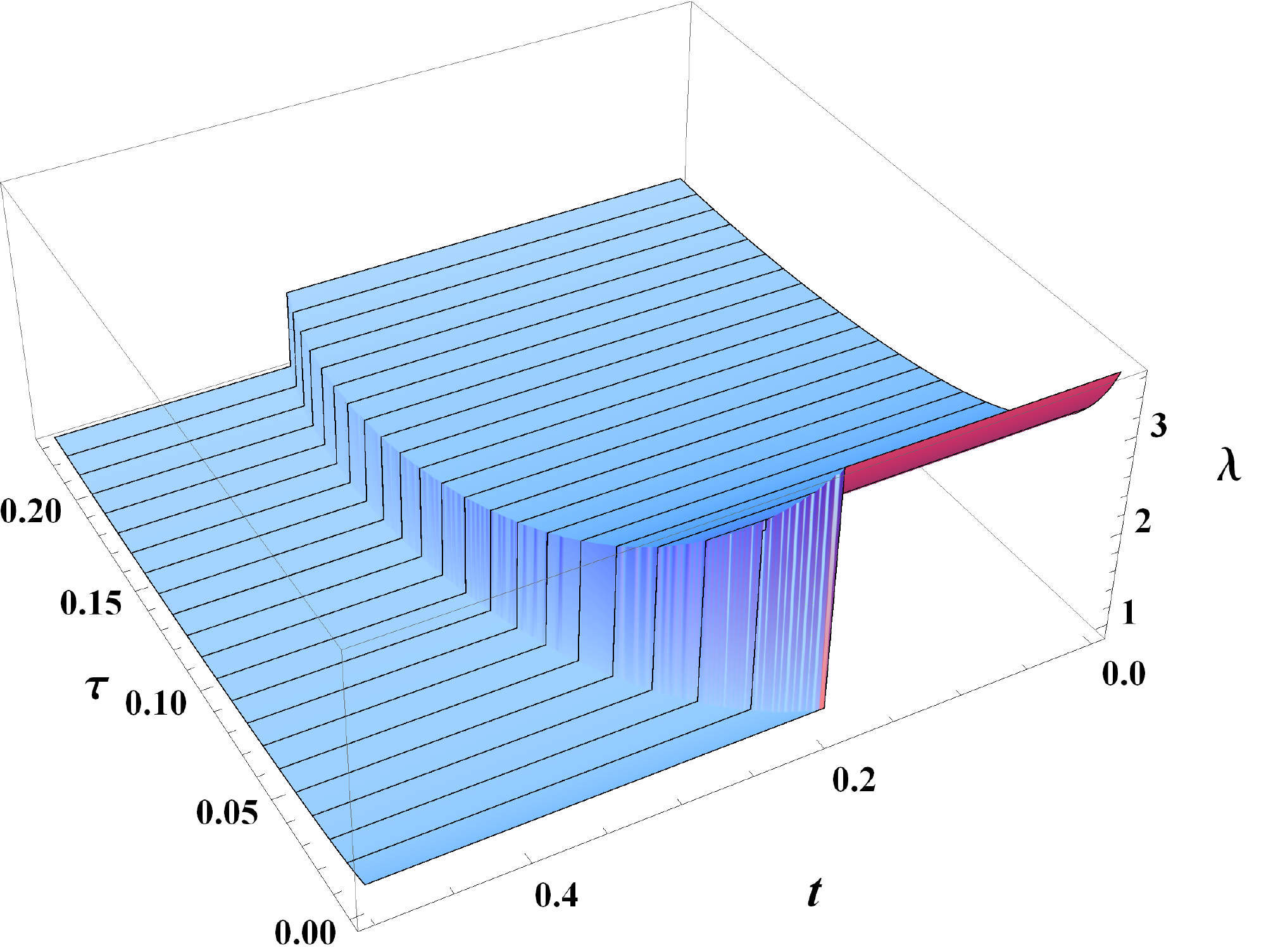}
    }
    \subfloat[d][\label{fig:twoGainsEN}Rising exponential shape]{
            \includegraphics[width=0.49\textwidth]{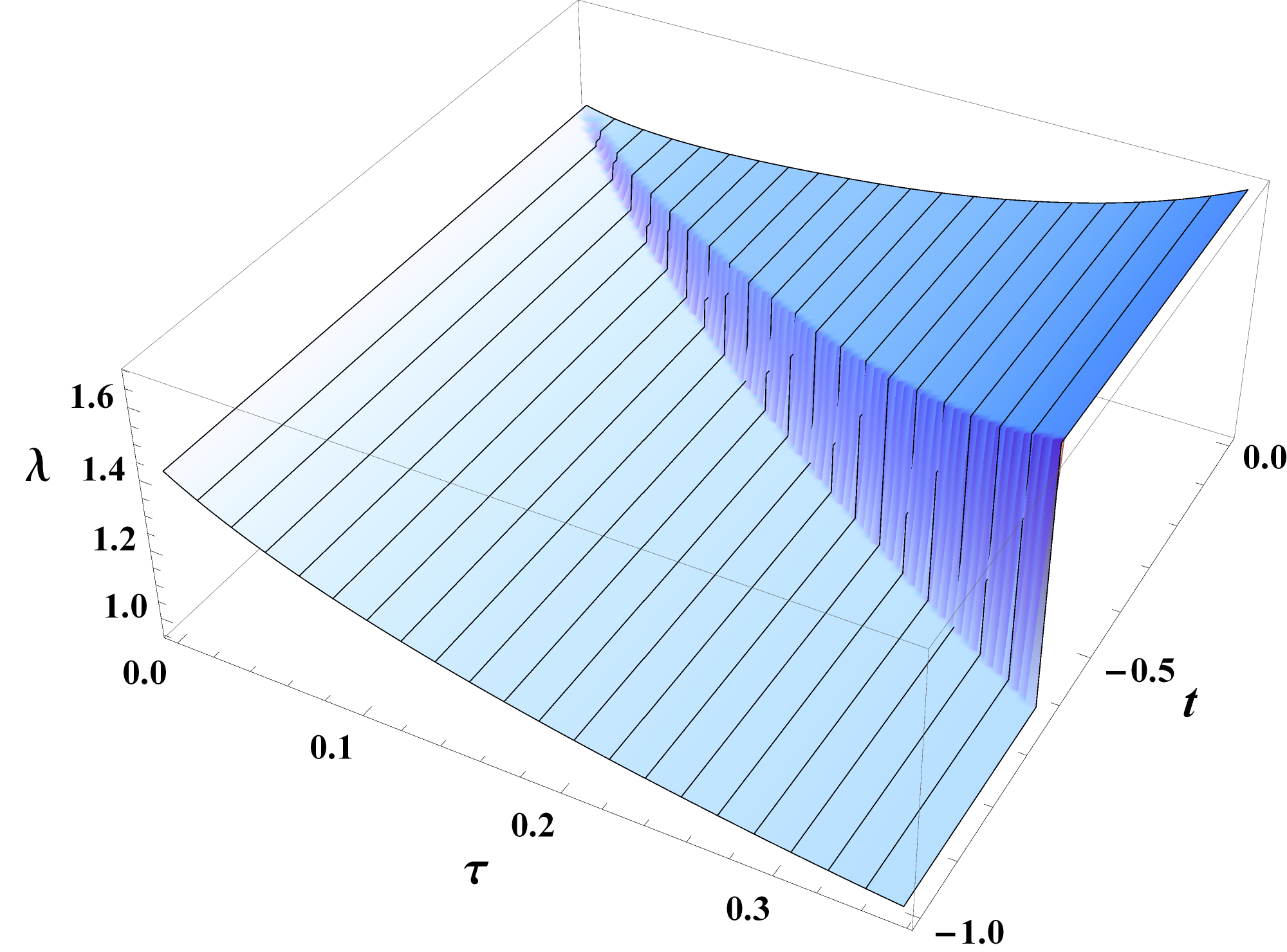}
    }
    \caption{\label{fig:twoGains}(Color online) Family of optimized piecewise constant feedback strategy $\y(t)$, plotted with respect to the delay $\delay$. Different mode-shapes has been considered: rectangular \eqref{fig:twoGainsR}, bilateral exponential \eqref{fig:twoGainsB}, falling  \eqref{fig:twoGainsEP} and  rising \eqref{fig:twoGainsEN} unilateral exponential. 
    }  
\end{figure*}
\begin{figure}
\centering
\includegraphics{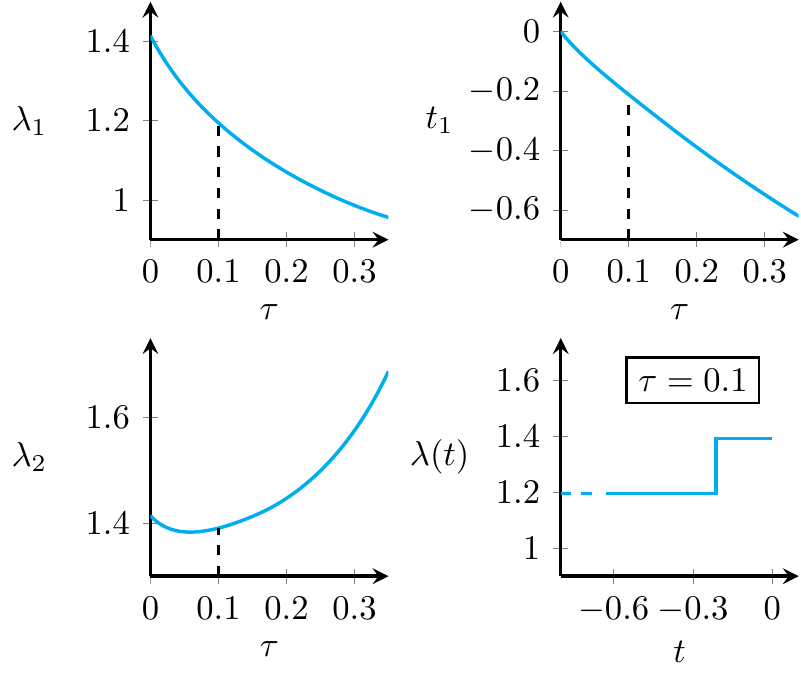}
\caption{\label{fig:seven}(Color online) Plots of the feedback parameters $\y_1$, $\y_2$ and $\tl$ of the optimized piecewise constant strategy $\y(t)$ as a function of the delay $\delay$ (upper left, lower left and upper right plot respectively) in the case of the rising exponential shape. In the lower right plot, the feedback gain $\y(t)$ is depicted for $\delay = 0.1$.
}
\end{figure}

In Fig. \ref{fig:twoGains} the family of the optimized piecewise constant feedback $\y(t)$ are plotted, as a function of the delay $\delay$. In each figure, black stripes has been plotted parallel to the time ($t$) axis, to indicate the shape of the function $\y(t)$ corresponding to a specific parameter $\delay$.  
 In all cases except the rising unilateral exponential, $\y(t)$ decreases with time; that is, $\y_1 \geq \y_2$.   In the case of the rectangular and bilateral exponential shapes (Figs. \ref{fig:twoGainsR} and \ref{fig:twoGainsB} respectively), $\y_2$ and $\tl$ shows almost no dependence, while $\y_1$ decreases with the delay. A similar behavior is exhibited in Fig. \ref{fig:twoGainsEP} for the falling unilateral exponential, with a little dependance of $\tl$ on the delay. By contrast, in the case of the rising unilateral exponential shape, Fig. \ref{fig:twoGainsEN}, $\y(t)$ increases with time; that is, $\y_1 \leq \y_2$. 
As the delay increases $\y_1$ and $\tl$ decrease, while $\y_2$ initially decreases but then increases. Fig. \ref{fig:seven} shows the same data as Fig. \ref{fig:twoGainsEN}, plotted as $\lambda_1$, $\lambda_2$ and $\tl$ versus $\delay$ for clarity. In the lower right plot, the optimal parameter for $\delay=0.1$ are collected to picture the optimal feedback gain $\y(t)$.

\section{Conclusion}
\label{sec:conc}

It has previously been shown possible to use adaptive homodyne detection to deterministically prepare an arbitrary single-rail qubit.  The protocol takes advantage of the result that a phase measurement on one half of an entangled state will collapse the other half into a known state, whose phase can be corrected if required. In this paper, we have studied that idea in detail from the experimental perspective.  We have defined a figure of merit for the adaptive homodyne state-preparation protocol and then investigated how various versions of the protocol perform.  

We have shown that ideal implementations of the protocol require the use of time-dependent gain in a real-time feedback loop in general.  Unfortunately a time-dependent gain in the feedback loop is difficult to implement experimentally, and a 
divergent gain -- as required for most mode-shapes when the single photon pulse first turns on -- 
 is impossible to implement experimentally. Also the ideal feedback algorithm assumes 
zero time delay in the feedback loop, which is unrealistic in general and particularly unrealistic in situations requiring implementation of complex, time-dependent feedback gain.  Therefore much of the paper was focused analysing the performance non-ideal, but experimentally feasible feedback algorithms in the adaptive homodyne state-preparation protocol.  In particular, we considered the performance of time-delayed constant or piecewise-constant feedback applied to a variety of different single-photon input wave-packets.   We have shown that these simple feedback algorithms can perform surprisingly well, thereby lending weight to the experimental viability of this scheme as a protocol for deterministically producing single-rail optical qubits.

\ack 

This work was supported by the Australian Research Council Centre of Excellence CE110001027. 
Nicola Dalla Pozza acknowledges partial support from the QUINTET and QUANTUM FUTURE grants of the University of Padova, and hospitality from the Centre for Quantum Dynamics at Griffith University, where part of this work was performed.  

\appendix

\section{Deriving $\am$ for adaptive dyne detection}
\label{sec:fidelDyne}

We derive a general expression for $\am$ as functional of the gain $\y(t)$ and the pulse shape $\u(t)$, 
first for zero delay, $\tau=0$, and then for $\tau > 0$.

\subsection{Zero delay} 

Consider for simplicity a rectangular pulse shape, which is zero everywhere except 
\beq
\u(t) = 1,\quad 0 \leq t \leq 1.
\eeq
so that we can restrict the integrals to the domain $[0,1]$.  
By the definition of \eqref{defA},  we have
\begin{align}
|\R|^2 & = \int_0^1 \dt \ \int_0^1 \ds \ e^{i \Xt-i \Xs} \ \Et \Es = \int_0^1 \dt \ \int_0^1 \ds \ \cos(\Xt-\Xs) \ \Et \Es  = 1 + 2 \expterm,
\end{align}
where 
\beq
\expterm \triangleq \int_0^1 \dt \int_0^{t_{-}} \ds \ \cos(\Xt-\Xs) \ \Et \Es .
\eeq
Note that $\ave{\expterm}$ must vanish since $\ave{|\R|^2} = 1$. 
In terms of $\expterm$, our  figure of merit \eqref{defam} is  thus 
\begin{align}
\am & =  1  - \xfrac{\ave{\expterm^2}}{2}.
\label{squareRootExpansion}
\end{align} 
The integrand function $\cos(\Xt-\Xs)$, as well as others that follow, depends on the stochastic process  
\beq
\Xt-\Xs=\int_s^{t_{-}} \E(v) \ \dv =\int_s^{t_{-}} \dW(v),
\eeq
that is non-anticipating in $t$, but correlated with $\dW(v),\ v \in [s,t)$. The increment $\Et$ is hence uncorrelated from the other integrand function, allowing one to factor the expectation
\begin{align}
\ave{\expterm} & = \ave{\int_0^1 \dt \int_0^{t_{-}} \ds \ \cos(\Xt-\Xs) \ \Et \Es} = \int_0^1 \dt \int_0^{t_{-}} \ds \ \ave{\cos(\Xt-\Xs) \ \Es}\ave{\Et} = 0 
\end{align}
by the stochastic \ito  calculus. As noted above, this result is necessary, so we have the first check on 
 the consistency of our approach. 

The term $\ave{\expterm^2}$ is
\begin{align}
\ave{\expterm^2}  = & \int_0^1 \dt \int_0^t \ds \int_0^1 \dv \int_0^v \dz \ \ave{\cos(\Xt-\Xs) \cos(\Xv-\Xu) \ \Et \Es \Ev \Ez} \label{eq1b} .
\end{align} \blk
On the range of integration where $t \neq v$, $\Et$ and $\Ev$ are uncorrelated, the expectation factorizes,
and hence these regions do not contribute to the integral. Equation \eqref{eq1b} thus reduces to 
\begin{align}
\ave{x^2} = & \int_0^1 \dt \int_0^{t_{-}}\ds \int_0^{t_{-}} \dz \ \ave{\cos(\Xt-\Xs) \cos(\Xt-\Xu) \ \Es \Ez}\ave{\Et^2}. \nonumber 
\end{align}
By again splitting the range of integration $[0,t) \times [0,t)$ in the variables $\ds,\ \dz$ into three regions, i.e. $s=z$, $s<z$ and $s>z$, and then using the symmetry of the integrand on the regions $s<z$ and $s>z$, we obtain two terms:
\beq 
\int_0^1 \dt \int_0^{t_{-}} \ds \int_0^{t_{-}} \dz \ \ave{\cos^2(\Xt-\Xs) \delta(s-z) \Es \Ez} \label{eq3} 
\eeq 
and 
\beq
 2\int_0^1 \dt \int_0^{t_{-}} \ds \int_0^{s_{-}} \dz 
 \ave{\cos(\Xt-\Xs) \cos(\Xt-\Xu) \ \Es \Ez}. \label{eq2b}
\eeq 

The functions $\cos^2(\Xt-\Xs)$, $\cos(\Xt-\Xs)$ and $\cos(\Xt-\Xu)$ are not uncorrelated
with $\Es$ and $\Ez$.  However, we can apply Novikov's formula \cite{Novikov1965}, which in our case reads
\beq
\ave{\Et {\cal F} } = \ave{\frac{\delta {\cal F}}{\delta \Et} }
\label{derivative}
\eeq
where ${\cal F}$ is an arbitrary differentiable functional of the stochastic process ${\boldsymbol \E}$. Applying the formula, 
we get for the term \eqref{eq3}
\begin{align}
& \int_0^1 \dt \int_0^{t_{-}} \ds \ \ave{\cos^2(\Xt-\Xs)} = \int_0^1 \dt \int_0^{t_{-}} \ds \ \frac{1+\ave{\cos(2\Xt-2\Xs)}}{2}
\end{align}
while for term \eqref{eq2b}, splitting the interval $[u,t)$ in $[u,s) \cup [s,t)$ in order to consider uncorrelated stochastic process $\Xt-\Xs$ and $\Xs-\Xu$, we get
\begin{align}
\ave{\cos(\Xt-\Xs)\cos(\Xt-\Xu)\Es \Ez} &  = \langle \y(z)\y(s) \sin(2\Xt-2\Xs)\sin(\Xs-\Xu)  \notag\\ 
& \quad -  \y(z)\y(s)\cos(2\Xt-2\Xs)\cos(\Xs-\Xu) \rangle \\
& =  \y(z)\y(s) \ave{\sin(2\Xt-2\Xs)} \ave{\sin(\Xs-\Xu)}  \notag\\ 
& \quad  - \y(z)\y(s)\ave{\cos(2\Xt-2\Xs)}\ave{\cos(\Xs-\Xu)}. \label{sincos2}
\end{align}
Finally, we can calculate the integrand function using their Taylor series and the moments 
for a Gaussian stochastic process of mean $0$ and standard deviation $\sigma$,
\beq
\ave{G^p} =
\begin{cases}
0 & \textrm{if $p$ is odd}\\
\sigma^p(p-1)!! & \textrm{if $p$ is even}
\end{cases}, 
\eeq
to obtain
\begin{align}
\ave{\sin(\X_i - \X_j)} & =0 \notag \\
\ave{\cos(\X_i - \X_j)} & = e^{-\xfrac{\textrm{Var}[\X_i - \X_j]}{2}} \label{sincos}\\
\ave{\cos(2\X_i - 2\X_j)} & = e^{-2\textrm{Var}[\X_i - \X_j]} \notag
\end{align}
where 
\beq
\textrm{Var}[\X_i - \X_j] = \int_j^i \y^2(\tau) \ \d \tau.
\eeq
These expressions are used to evaluate \eqref{sincos2}, \eqref{eq3} and \eqref{eq2b}. Altogether, the final expression for \eqref{eq1b} is 
\begin{align}
\ave{\expterm^2} &=  \int_0^1 \dt \int_0^{t_{-}} \ds \ \rnfrac{1+e^{\textstyle -2\int_s^t \ytau^2 \ \d \tau}}{2} \nonumber \\
& \qquad - 2 \int_0^1 \dt \int_0^{t_{-}} \ds \int_0^{s_{-}} \dz \ \ys \yu e^{\textstyle -2\int_s^t \ytau^2 \ \d \tau} {e}^{\textstyle -\int_v^s \ytau^2 \ \d \tau/2}  ,
\end{align}
and by the definition \eqref{squareRootExpansion} we obtain
\begin{align}
\am &=  1 - \int_0^1 \dt \int_0^{t_{-}} \ds \ \rnfrac{1+e^{\textstyle -2\int_s^t \ytau^2 \ \d \tau}}{4} \nonumber \\
& \qquad + \int_0^1 \dt \int_0^{t_{-}} \ds \int_0^{s_{-}} \dz \ \ys \yu e^{\textstyle -2\int_s^t \ytau^2 \ \d \tau} {e}^{\textstyle -\int_v^s \ytau^2 \ \d \tau/2}.
\label{final}
\end{align} 
The generalization of Eq.\eqref{final} for an arbitrary mode-shape $\u(t)$ leads to the expression \eqref{final2}.

\subsection{Non-zero delay \label{sec:delay}}

In this section we generalize \eqref{final2} for a delay $\delay$ in the feedback scheme, 
by defining 
\begin{align}
\X(t) & = \frac{\pi}{2} + \int_{-\infty}^{t-\delay} \y(s) \  \dW(s).
\label{def:Xdelay}
\end{align}
In this case, $\expterm$ can be written as
\beq
\expterm = \int_{-\infty}^{+\infty} \dt \int_0^{+\infty} \dD \cos(\Xt-\XtD) \sqrt{\ut \utD} \Et \EtD.
\eeq 
As before, the term $\ave{\expterm}$ vanishes,
while after some algebra the term $\ave{\expterm^2}$ becomes  
\begin{align}
\ave{\expterm^2} & =  \int_{-\infty}^{+\infty} \dt \int_{0}^{+\infty} \dD \ \ave{\cos^2(\Xt-\XtD)\ut \utD} \notag \\
& \quad -4 \int_{-\infty}^{+\infty}\dt \int_\delay^{+\infty} \dD \int_0^\delay \dd \ \ytD \ytDd \ut \sqrt{\utD \utDd} \notag \\
& \qquad \qquad \times \ave{\cos(2\Xt-2\XtD) \cos(\XtD-\XtDd)} \label{eq5} \\
& \quad -2 \int_{-\infty}^{+\infty} \dt \int_\delay^{+\infty} \dD \int_\delay^{+\infty} \dd \ \ytD \ytDd \ut \sqrt{\utD \utDd} \notag \\
& \qquad \qquad \times \ave{\cos(2\Xt-2\XtD) \cos(\XtD-\XtDd)} \notag
\end{align}
The three terms in the integral comes from splitting the range of integration opportunely, in order to identify regions where only a few of the stochastic process $\Xt-\XtD,\ \Xt-\XtDd,\ \EtD,\ \Etd$ are correlated. Then, we use the rule \eqref{derivative} to get rid of the $\EtD,\ \Etd$. Finally, we resolve the expectation employing \eqref{sincos}, to 
obtain \eqref{final3}.

\section{Approximate Figure of Merit $\am$ for Constant Gain Feedback}
\label{AppendixB}

In this appendix we report the analytical expressions evaluated from Eq. \eqref{final3} employing a constant gain feedback and the mode shapes introduced in Section \ref{sec:mode-shapes}. These closed form solutions allow to easily set up a numerical optimization of the feedback gain and maximize the figure of merit $\am$. The result of the optimization has been discussed in Section \ref{sec:constantGain} and are depicted in Fig.~\ref{fig:constantGainPerformances} and \ref{fig:constantGain}.

In the case of the rectangular mode-shape, we have different expressions depending upon the value of the delay. We are interested in the range $0<\delay<T/2$, since for larger delays the adaptive dyne scheme does not lead to any improvements with respect to the heterodyne measurement. With these assumptions, the expression for $\am$  reduces to Eq. \eqref{rectY}.

In the cases of the other mode shapes considered, the range of $\delay$ is not restricted to any interval. Employing the bilateral exponential mode-shape, we obtain the expression \eqref{bilateralY}. Instead, in the case of falling exponential mode-shape, the expression reduces to
\beq
\am = 1-\frac{3k^2+4\left [1-2{e}^{-( k+2\y^2)\delay+4{e}^{-\frac{5}{2}(k+\y^2)\delay}}\right ]k \y^2+ \y^4}{4(3k^2+7k\y^2+2\y^4)}, 
\eeq
while substituting the rising exponential mode-shape, we get
\beq
\am = 1-\frac{k^2+2\left [ 1-4{e}^{-(k+2\y^2)\delay}+2{e}^{-\frac{1}{2}(3 k+5\y^2)\delay} \right ] k \y^2+ \y^4}{4(k^2+3k\y^2+2\y^4)}.
\eeq
 
\begin{align}
& \am = 1-\frac{1}{16} \left [2 T \left(T+\frac{1}{\y^2}\right)+\frac{e^{-2 T \y^2}-1}{\y^4}\right ] - \frac{e^{-2 \y ^2 \tau } \left(-2 T \y ^2+2 \y ^2 \tau +5\right)}{\y ^4} \notag \\
& \qquad - \frac{e^{-2 T \y ^2} \left(2-e^{\frac{3 \y ^2 \tau }{2}}\right)}{6 \y ^4}
+\frac{8 e^{-\frac{1}{2} \y ^2 (T+3 \tau )}}{3 \y ^4}
+\frac{e^{-\frac{5 \y ^2 \tau }{2}} \left(-2 T \y ^2+4 \y ^2 \tau +5\right)}{2 \y ^4}
\label{rectY}
\end{align}

\begin{align}
& \am =1-\frac{1}{8 \left(k+2 \y^2\right)^2 \left(3 k^2+4 k\y^2+\y^4\right)} \notag\\
& \quad \times \bigg \{ 6 k^4+23 k^3 \y^2+34 k^2 \y^4+21 k \y^6+4 \y^8 -2 k \y^2 \left(k^2+3 k \y^2+2 \y^4\right) e^{-\frac{5}{2} \tau  \left(k+\y^2\right)} \notag \\
& \quad \qquad -8 k \y^2  \left(3 k^3 \tau +7 k^2 \left(\y^2 \tau +1\right)+2 k \y^2 \left(\y^2 \tau +5\right)+2 \y^4\right) e^{ - \tau  \left(k+2 \y^2\right)} \notag\\
& \quad \qquad 	2 k \y^2 \left(3 k+\y^2\right)\left(2 k^2 \tau +k \left(4 \y^2 \tau +5\right)+6 \y^2\right)  e^{-\frac{1}{2} \tau \left(3 k+5 \y^2\right)} 
\bigg \}\label{bilateralY}
\end{align}

\section*{References}

\bibliography{qubitStatePreparation}

\end{document}